\documentclass[aps,prc,preprint,groupedaddress,showpacs]{revtex4}
\usepackage[dvipdfm]{graphicx}
\usepackage{amsmath}

\begin{document}
\title{Two-phonon $\gamma$-vibrational states in rotating triaxial odd-$A$ nuclei}

\author{Masayuki Matsuzaki}
\email[]{matsuza@fukuoka-edu.ac.jp}
\affiliation{Department of Physics, Fukuoka University of Education, 
             Munakata, Fukuoka 811-4192, Japan}

\date{\today}

\begin{abstract}
Distribution of the two phonon $\gamma$ vibrational collectivity 
in the rotating triaxial odd-$A$ nucleus, $^{103}$Nb, 
that is one of the three nuclides for which experimental data were 
reported recently, is calculated in the framework of the particle 
vibration coupling model based on the cranked shell model plus 
random phase approximation. This framework was previously utilized 
for analyses of the zero and one phonon bands in other mass region 
and is applied to the two phonon band for the first time. 
In the present calculation, three sequences 
of two phonon bands share collectivity almost equally at finite 
rotation whereas the $K=\Omega+4$ state is the purest at zero rotation. 
\end{abstract}

\maketitle

\section{Introduction}

One of the properties most specific to the finite quantum many body 
system, the atomic nucleus, is to exhibit both single particle and 
collective modes of excitation in similar energy scale. 

Thanks to recent progress in computer power, exact diagonalization 
of given effective interaction in a huge model space is becoming 
available for up to medium mass nuclei including those far from 
stability. But in order to extract physical picture of their dynamics 
from huge numerical information, it is necessary to have recourse 
to the concept of collectivity. Among the various kinds of collective 
modes of excitation, one of those that have been well studied and 
known to exist prevailingly in the nuclear chart is the $\gamma$ 
vibration. Vibrational excitations in many fermion systems, in other 
terms, phonons, are made up by coherent superposition of 
particle-hole or two quasiparticle excitations. Thus, on the one 
hand, multiple excitations are expected if they are really collective. 
On the other hand, their excitation spectra, in particular their 
anharmonicity reflect underlying nuclear structure. 

In the nuclear chart, collective vibrations in the rare earth nuclei 
with the mass $A=$160 - 170 are the best studied. The first experimental 
information about the two phonon $\gamma$ vibration, denoted by the 
2$\gamma$ hereafter, in $^{168}$Er was reported by Davidson et 
al.~\cite{Dav} based on a high resolution $\gamma$ ray study following 
neutron capture. The first theoretical analysis was done by Warner et 
al.~\cite{WCS} in terms of the interacting boson model with $s$ and 
$d$ bosons but it was critically assessed by Bohr and 
Mottelson~\cite{BMPS}. 
They argued within the general framework applicable to deformed nuclei 
with rotational spectra~\cite{BM}. Dumitrescu and Hamamoto~\cite{DH} 
elucidated this problem by means of a macroscopic and microscopic analysis. 
On the contrary, Soloviev and Shirikova~\cite{SS} argued that collective 
two phonon excitations do not exist because of the strong Pauli principle 
among nucleons that constitute vibrational excitations based on the 
quasiparticle phonon model. Matsuo~\cite{Matsuo} and Matsuo and 
Matsuyanagi~\cite{MatsuoMatsu} applied the selfconsistent collective 
coordinate method that provides quantized collective Hamiltonian starting 
from the microscopic random phase approximation (RPA) and obtained the 
result that the $K=4$ 2$\gamma$ state in the triaxially deformed 
potential exists at the energy about 2.7 times that of the 1$\gamma$ 
state. Pipenbring and Jammari~\cite{PJ} also obtained a similar result 
by means of the multiphonon method based on the Tamm-Dancoff approximation. 
The description in terms of the 
interacting boson model was improved by Yoshinaga et al.~\cite{YAA} 
by introducing the $g$ boson. 
The definite experimental evidence, the absolute $B(E2)$ value between 
the $K=4$ 2$\gamma$ candidate and the 1$\gamma$, that proves that the 
$K=4$ state is really the 2$\gamma$ was given by means of $\gamma$ ray 
induced Doppler broadening following neutron capture by B\"{o}rner et 
al.~\cite{Bo} and using Coulomb excitation by Oshima et al.~\cite{Oh} 
and H\"{a}rtlein et al.~\cite{HHSF}. 

 The 2$\gamma$ states of very similar character were 
predicted~\cite{MatsuoMatsu,SSS} and observed in nearby 
nuclides, $^{166}$Er using Coulomb excitation by Fahlander et 
al.~\cite{Fa} and using a $(n,n'\gamma)$ reaction by Garrett et 
al.~\cite{Ga}, and $^{164}$Dy using thermal neutron capture by 
Corminboeuf et al.~\cite{Co}. In particular, the $K=0$ 2$\gamma$ 
state was also observed in $^{166}$Er~\cite{Ga}. 
Sun et al.~\cite{Sun} studied the 2$\gamma$ bands in $^{166,168}$Er
in terms of the triaxial projected shell model. This model gives the 
$K=4$ 2$\gamma$ states in a kinematical manner similar to 
Davydov-Filippov's asymmetric rotor model~\cite{DF}. 
In other mass regions, 
harmonic 2$\gamma$ states were observed in $^{232}$Th by Korten et 
al.~\cite{Kor1,Kor2} using Coulomb excitation 
and $^{106,104}$Mo by Guessous et al.~\cite{Mo106,Mo104} using 
a spontaneous fission. The latter was discussed from a viewpoint of 
the $X(5)$ symmetry~\cite{BB}. These very limited number of observations 
indicate that it depends strongly on the underlying single particle 
level structure whether the 2$\gamma$ states exist or not. This fact 
suggests that microscopic description of the collective vibrational 
excitations is mandatory. 

In odd-$A$ nuclei, some numerical prediction for rare earth nuclides 
were made by Durand and Piepenbring~\cite{DP} in terms of the 
multiphonon method prior to experimental observation. The first 
experimental observation was made ten years later in $^{105}$Mo 
by Ding et al.~\cite{Mo105} using a spontaneous fission. In these 
fission fragments, $^{104 - 106}$Mo, rotational band members built on the 
2$\gamma$ states were also populated. Soon after this, similar 
rotational bands were observed also in $^{103}$Nb by Wang et 
al.~\cite{Wa} and $^{107}$Tc by Long et al.~\cite{Lo}. These nuclei 
exhibit anharmonicity $E_{2\gamma}/E_{1\gamma}<2$ in $^{105}$Mo and 
$^{103}$Nb, while $\agt 2$ in $^{107}$Tc, where level energies are 
measured from the corresponding zero phonon states. These are the 
2$\gamma$ states observed so far. 
The first theoretical calculation to the observed 2$\gamma$ band in 
the odd-$A$ nucleus, $^{103}$Nb, was done by Sheikh et al.~\cite{SBSP} 
in terms of the triaxial projected shell model. It was discussed there 
that the observed anharmonicity is difficult to be reproduced with the 
triaxial parameter that gives a good description of the 1$\gamma$ band. 

In the present paper, we take a complementary approach to the 
2$\gamma$ band in $^{103}$Nb; the particle vibration coupling (PVC) 
calculation based on the RPA phonons constructed in the rotating 
frame allowing possible static triaxial deformation. 

\section{The model}

In the model adopted in this study, elementary modes are quasiparticles 
and $\gamma$ vibrational RPA phonons excited on top of common 
vacuum as in traditional calculations, however, the vacuum is the 
yrast configuration of an even-even nucleus rotating with a frequency 
$\omega_\mathrm{rot}$. This vacuum configuration can be either zero 
quasiparticle, two quasiparticle and so on, seen from the non-rotating 
ground state. Excitations are labeled by the signature quantum number 
$r=\exp{(-i\pi\alpha)}$, $I=\alpha+\mathrm{even}$, 
appropriate for rotating reflection-symmetric 
objects. Two kinds of $\gamma$ vibrational excitations exist, denoted 
by $X_{\gamma(\pm)}^\dagger$ with $r=\pm$1, respectively. The phonon 
space is limited to zero, one and two phonon states. The vacuum 
mean field is rotating and static triaxial deformation is also possible; 
this means that the $\Omega$-mixing in quasiparticles and the $K$-mixing 
in RPA phonons are naturally taken into account. Here $\Omega$ is the 
projection of the single particle angular momentum to the third axis. 

The formulation is summarized as follows. 
We begin with a one-body Hamiltonian in the rotating frame, 
\begin{gather}
h'=h-\omega_\mathrm{rot}J_x , \\
h=h_\mathrm{Nil}-\Delta_\tau (P_\tau^\dagger+P_\tau)
                   -\lambda_\tau N_\tau , \label{hsp} \\
h_\mathrm{Nil}=\frac{\mathbf{p}^2}{2M}
                +\frac{1}{2}M(\omega_x^2 x^2 + \omega_y^2 y^2 + \omega_z^2 z^2)
                +v_{ls} \mathbf{l\cdot s} 
                +v_{ll} (\mathbf{l}^2 - \langle\mathbf{l}^2\rangle_{N_\mathrm{osc}}) .
                \label{hnil}
\end{gather}
In Eq.~(\ref{hsp}), $P_\tau$ is the pair annihilation operator, 
$\tau = 1$ and 2 denote neutron and proton, respectively, 
and the chemical potentials $\lambda_\tau$ are determined so as to give the 
correct average particle numbers $\langle N_\tau \rangle$. 
The oscillator frequencies in Eq.~(\ref{hnil}) 
are related to the quadrupole deformation parameters $\epsilon_2$ and $\gamma$ 
in the usual way. 
They, along with the pairing gaps $\Delta_\tau$, are 
determined from experimental information. 
The orbital angular momentum $\mathbf{l}$ in Eq.~(\ref{hnil}) is defined in 
the singly stretched coordinates $x_k' = \sqrt{\frac{\omega_k}{\omega_0}}x_k$ 
and the corresponding momenta, with $k =$ 1, 2 and 3 denoting $x$, $y$ and $z$, 
respectively. We apply the RPA to the residual pairing plus 
doubly stretched quadrupole-quadrupole ($Q'' \cdot Q''$) interaction between 
quasiparticles. It is given by  
\begin{equation}
H_\mathrm{int}=
-\sum_{\tau=1,2} G_\tau \tilde P_\tau^\dagger \tilde P_\tau
-\frac{1}{2}\sum_{K=0,1,2} \kappa_K^{(+)} Q_K''^{(+)\dagger} Q_K''^{(+)} 
-\frac{1}{2}\sum_{K=1,2} \kappa_K^{(-)} Q_K''^{(-)\dagger} Q_K''^{(-)} ,
\end{equation}
where the doubly stretched quadrupole operators are defined by 
\begin{equation}
Q_K''=Q_K(x_k\rightarrow x_k'' = \frac{\omega_k}{\omega_0}x_k) ,
\end{equation}
and those with good signature are 
\begin{equation}
Q_K^{(\pm)}=\frac{1}{\sqrt{2(1+\delta_{K0})}}\left(Q_K \pm Q_{-K}\right) ,
\end{equation}
and $\tilde P_\tau$ is defined by subtracting the vacuum expectation value 
from $P_\tau$. 
Among RPA modes determined by the equation of motion, 
\begin{equation}
\left[h'+H_\mathrm{int},X_n^\dagger\right]_\mathrm{RPA}
=\omega_n X_n^\dagger ,
\end{equation}
we choose the $\gamma$ vibrational phonons, $n=\gamma(\pm)$, which have 
outstandingly large $K=2$ transition amplitudes
\begin{equation}
\bigl|T_K^{(\pm)}\bigr|
=\Bigl|\left\langle\left[Q_K^{(\pm)},X_{\gamma(\pm)}^\dagger\right]
 \right\rangle\Bigr| .
\end{equation}

The particle vibration coupling Hamiltonian takes the form
\begin{equation}
\begin{split}
H_\mathrm{couple}(\gamma)
     &={\sum_{\mu\nu}}\Lambda_{\gamma(+)}(\mu\nu)
       \left(X_{\gamma(+)}^\dagger a_\mu^\dagger a_\nu
           + X_{\gamma(+)}         a_\nu^\dagger a_\mu\right) \\
     &+{\sum_{\bar\mu\bar\nu}}\Lambda_{\gamma(+)}(\bar\mu\bar\nu)
       \left(X_{\gamma(+)}^\dagger a_{\bar\mu}^\dagger a_{\bar\nu}
           + X_{\gamma(+)}         a_{\bar\nu}^\dagger a_{\bar\mu}\right) \\
     &+\sum_{\mu\bar\nu}\Lambda_{\gamma(-)}(\mu\bar\nu)
       \left(X_{\gamma(-)}^\dagger a_\mu^\dagger a_{\bar\nu}
           + X_{\gamma(-)}         a_{\bar\nu}^\dagger a_\mu\right) \\
     &+\sum_{\mu\bar\nu}\Lambda_{\gamma(-)}(\bar\nu\mu) 
       \left(X_{\gamma(-)}^\dagger a_{\bar\nu}^\dagger a_\mu
           + X_{\gamma(-)}         a_\mu^\dagger a_{\bar\nu}\right) ,
\end{split}
\end{equation}
where $\mu$ and $\bar\mu$ denote quasiparticles with $r=-i$ and $+i$, 
respectively. The coupling vertices are given by
\begin{equation}
\begin{split}
   &\Lambda_{\gamma(+)}(\mu\nu)
             =-\sum_{K=0,1,2}\kappa_K^{(+)} T_K''^{(+)}
                             Q_K''^{(+)}(\mu\nu) ,\\
   &\Lambda_{\gamma(+)}(\bar\mu\bar\nu)
             =-\sum_{K=0,1,2}\kappa_K^{(+)} T_K''^{(+)}
                             Q_K''^{(+)}(\bar\mu\bar\nu) ,\\
   &\Lambda_{\gamma(-)}(\mu\bar\nu)
             =-\sum_{K=1,2}\kappa_K^{(-)} T_K''^{(-)}
                             Q_K''^{(-)}(\mu\bar\nu) ,\\
   &\Lambda_{\gamma(-)}(\bar\nu\mu)
             =-\sum_{K=1,2}\kappa_K^{(-)} T_K''^{(-)}
                             Q_K''^{(-)}(\bar\nu\mu) ,
\end{split}
\end{equation}
where $Q_K''^{(\pm)}(\alpha\beta)$ denotes quasiparticle scattering matrix 
elements that do not contribute to RPA phonons. Eigenstates of the 
Hamiltonian thus specified at each $\omega_\mathrm{rot}$ take the form 
\begin{equation}
\begin{split}
\left.|\chi_j\right\rangle 
&=\sum_{\mu}\psi_j^{(1)}(\mu)\left.a_\mu^\dagger|\phi\right\rangle \\
& +\sum_{\mu}\psi_j^{(3)}(\mu\gamma)\left.a_\mu^\dagger X_\gamma^\dagger|\phi\right\rangle
  +\sum_{\bar\mu}\psi_j^{(3)}(\bar\mu\bar\gamma)
\left.a_{\bar\mu}^\dagger X_{\bar\gamma}^\dagger|\phi\right\rangle \\
& +\sum_{\mu}\psi_j^{(5)}(\mu\gamma\gamma)\frac{1}{\sqrt{2}}
\left.a_\mu^\dagger X_\gamma^\dagger X_\gamma^\dagger|\phi\right\rangle
  +\sum_{\mu}\psi_j^{(5)}(\mu\bar\gamma\bar\gamma)\frac{1}{\sqrt{2}}
\left.a_\mu^\dagger X_{\bar\gamma}^\dagger X_{\bar\gamma}^\dagger|\phi\right\rangle \\
& +\sum_{\bar\mu}\psi_j^{(5)}(\bar\mu\gamma\bar\gamma)
\left.a_{\bar\mu}^\dagger X_\gamma^\dagger X_{\bar\gamma}^\dagger|\phi\right\rangle , \\
&\mbox{for the $r=-i$ sector} ,
\end{split}
\end{equation}
where $\gamma$ and $\bar\gamma$ abbreviate $\gamma(+)$ and $\gamma(-)$, 
respectively, and $\left.|\phi\right\rangle$ is the rotating vacuum 
configuration. Those for the $r=+i$ sector take a form similar to above, 
except that the suffices $\mu$ are to be replaced by $\bar\mu$. 

This model was first developed for studying the signature dependence 
of the level energies and $E2$, $M1$ transition rates in one quasiparticle 
(zero phonon) bands~\cite{MSM,MM}, and then applied to study the $E2$ 
intensity relation~\cite{BM,SN} in the 1$\gamma$ rotational bands~\cite{Ge}. 
The present study is the first application to the 2$\gamma$ states in 
rotating odd-$A$ nuclei. By construction of the model space, this model is 
applicable up to the energy region $\alt E_\mathrm{1qp}+2\Delta$; otherwise 
non-collective 3qp states dominate over the collective states. 

\section{Results and discussions}

\begin{figure}[htbp]
 \includegraphics[width=8cm,angle=-90]{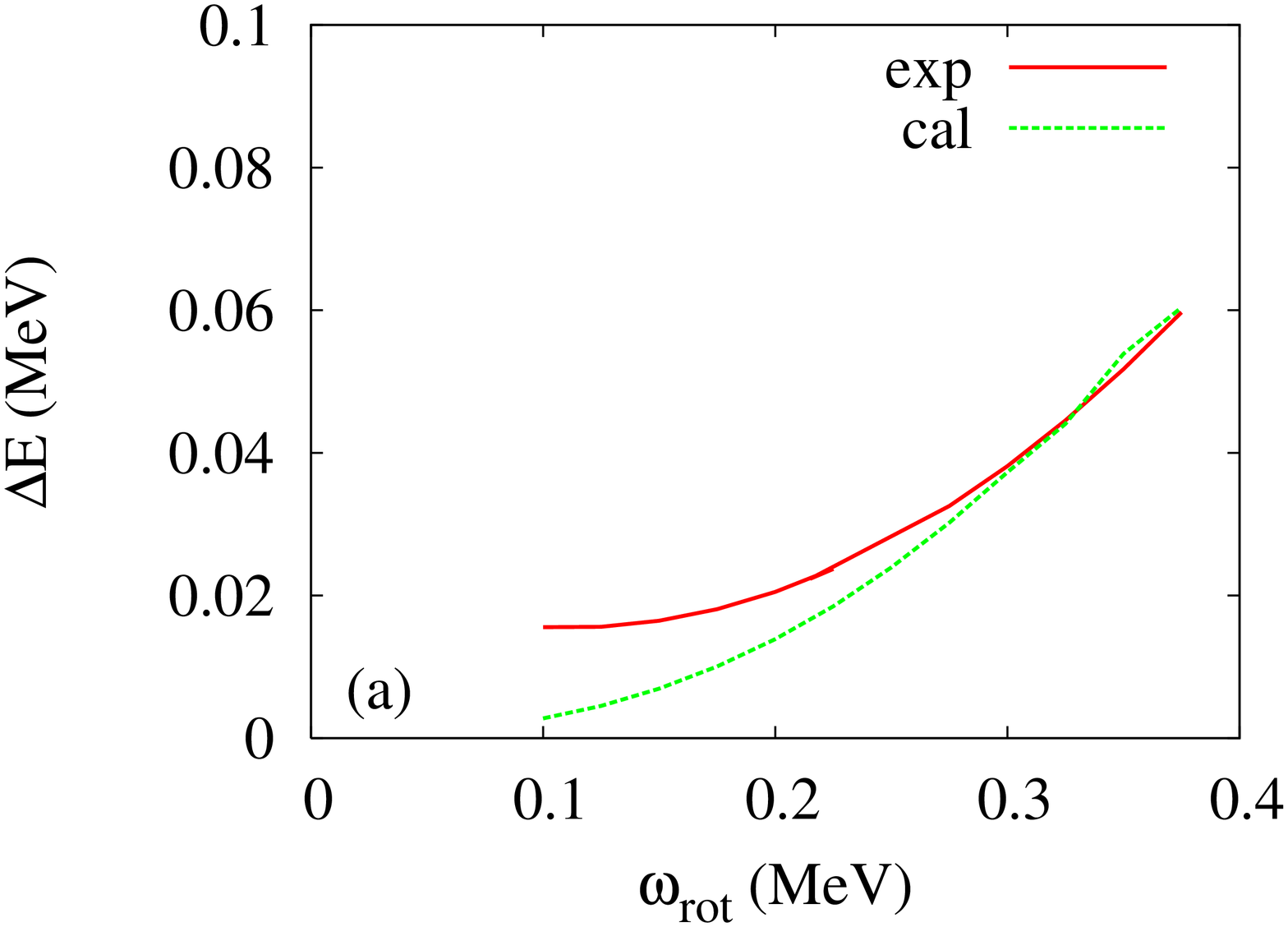}
 \includegraphics[width=8cm,angle=-90]{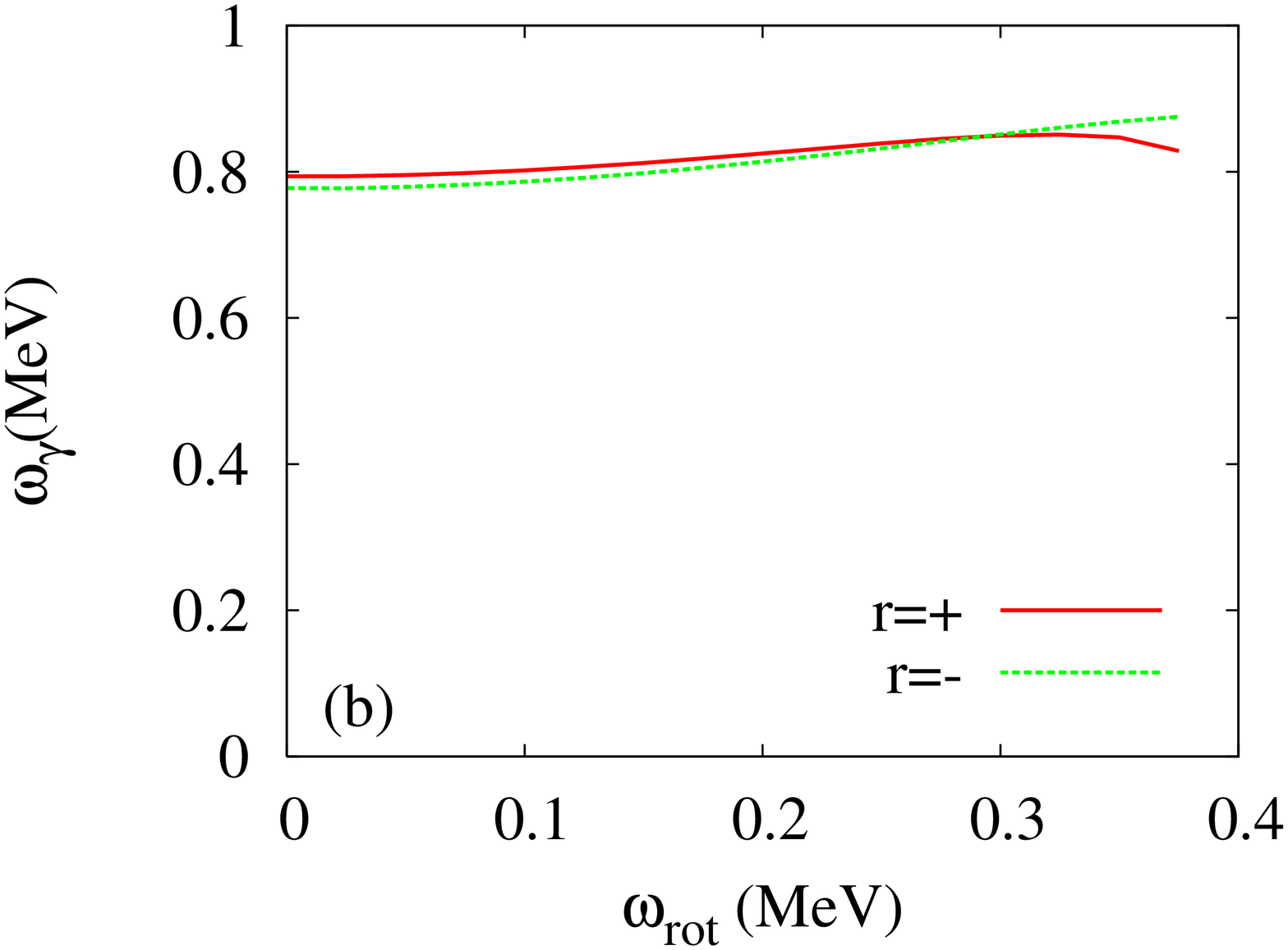}
 \caption{(Color online) (a) Experimental and calculated signature splitting 
in the $\pi[422\,5/2^+]$ one quasiparticle band as functions of the rotation 
frequency. Theoretical curve is the result of the particle vibration coupling 
rather than the cranked shell model alone. The latter (not shown) is larger 
than the former. (b) Excitation energies of the two types of $\gamma$ 
vibrational RPA phonons as functions of the rotation frequency.}
 \label{fig1}
\end{figure}

Three new rotational bands that feed the $\pi[422\,5/2^+]$ ground 
band of $^{103}_{41}$Nb$_{62}$ were observed by a very recent $\gamma$ ray 
study of spontaneous fission fragments from $^{252}$Cf~\cite{Wa}. 
Among them, the one that is built on the $9/2^+$ bandhead was assigned 
to the $K=\Omega+2$ sequence of the 1$\gamma$ bands. One of the others 
that is built on the $13/2^+$ bandhead was assigned to the $K=\Omega+4$ 
sequence of the 2$\gamma$ bands. This is the second observation of the 
2$\gamma$ band in odd-$A$ nuclei, and to which the first theoretical 
calculation~\cite{SBSP} was reported. In the present cranked shell model 
calculation, diagonalization is performed in the five major 
shells, $N_\mathrm{osc}=$ 2 - 6 for the neutron and 1 - 5 for the proton 
with the Nilsson parameters $v_{ls}$ and $v_{ll}$ taken from 
Ref.~\cite{BR}. The three mean field parameters, which are assumed to be 
$\omega_\mathrm{rot}$ independent for simplicity, the pairing gaps 
$\Delta_n=$ 1.05 MeV, $\Delta_p=$ 0.85 MeV and the deformation 
$\epsilon_2=$ 0.31 are adopted from the experimental analyses~\cite{Mo106,Wa}. 
The triaxiality $\gamma$ is chosen so as to reproduce the measured signature 
splitting of the ground band in the PVC calculation. The chosen value, 
$\gamma=-7^\circ$, gives overall reproduction of the signature splitting 
in the rotating frame aside from near the bandhead as shown in 
Fig.~\ref{fig1}(a).
Note here that it was reported that the signature splitting 
can be reproduced without invoking $\gamma$ deformation in the 
ancestral model~\cite{HS} of the triaxial projected shell model. 
This suggests that the appropriate value of the 
$\gamma$ deformation is model dependent as noticed in Ref.~\cite{GCS} 
and discussed below. 

The strengths of the residual doubly stretched quadrupole interaction are 
determined as follows; in the reference configuration with 
$\omega_\mathrm{rot}=0$ and $\gamma=0$ in which $K$ is a good quantum number, 
$\kappa_2^{(+)}=\kappa_2^{(-)}$ is determined to reproduce within the RPA 
the observed $\gamma$ vibrational energy in the adjacent $^{104}$Mo, 0.812 
MeV. If fully collective $\beta$ vibration exists, $\kappa_0^{(+)}$ 
can be determined to reproduce its energy but since the collective character 
of the observed $0^+_2$ is not clear, $\kappa_0^{(+)}$ is set equal to 
$\kappa_2^{(\pm)}$. And $\kappa_1^{(+)}=\kappa_1^{(-)}$ is determined so as to 
make the energy of the Nambu-Goldstone mode zero. Those of the residual 
pairing interaction are determined to reproduce the adopted pairing gaps. 
Then the RPA calculation is performed with $\gamma=-7^\circ$. The obtained 
$\omega_\mathrm{rot}$ dependence and signature splitting~\cite{BM,SM} of the 
excitation energy of $\gamma$ vibration is weak as shown in Fig.~\ref{fig1}(b).

\begin{figure}[htbp]
 \includegraphics[width=5cm,angle=-90]{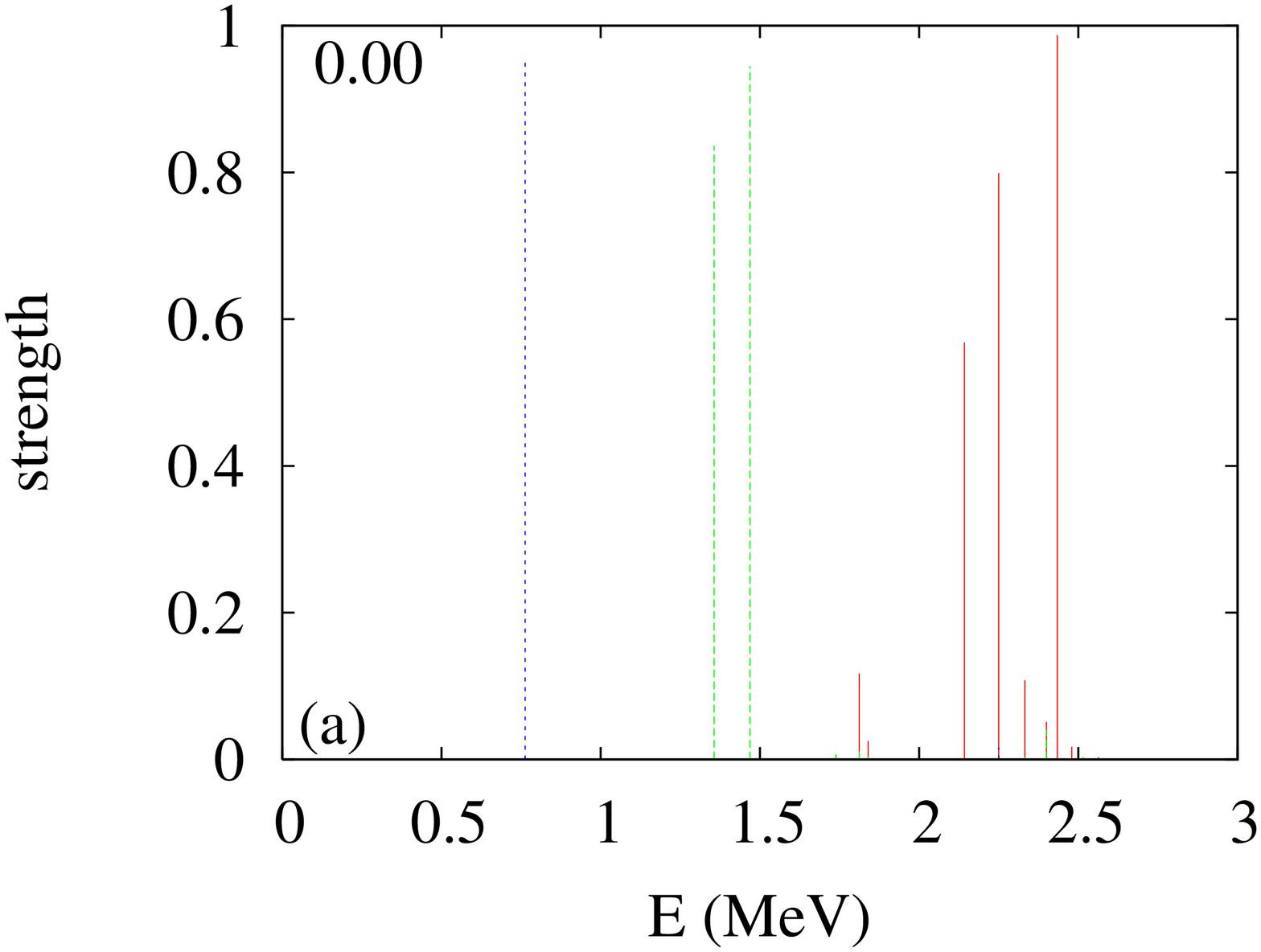}
 \includegraphics[width=5cm,angle=-90]{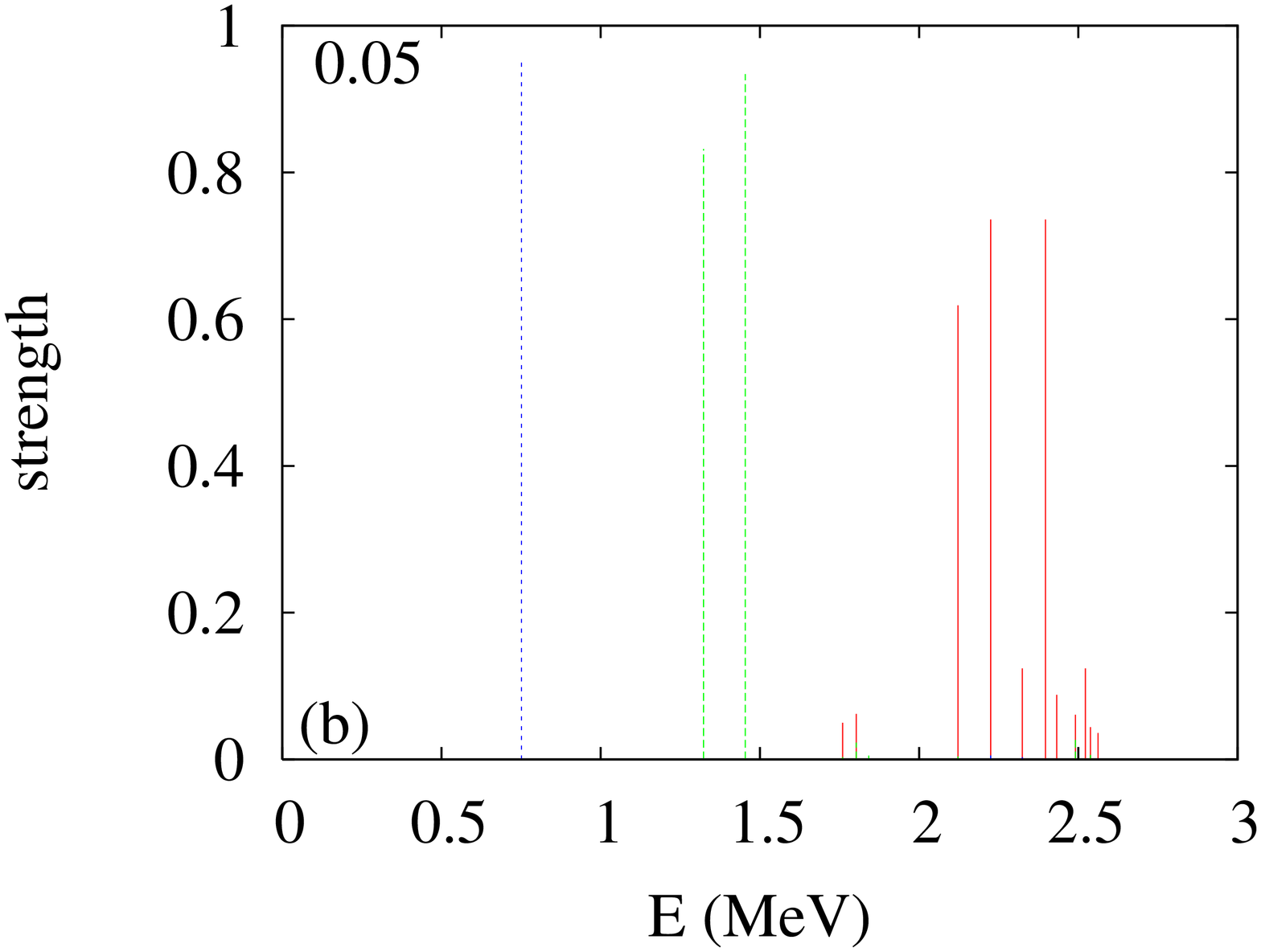}
 \includegraphics[width=5cm,angle=-90]{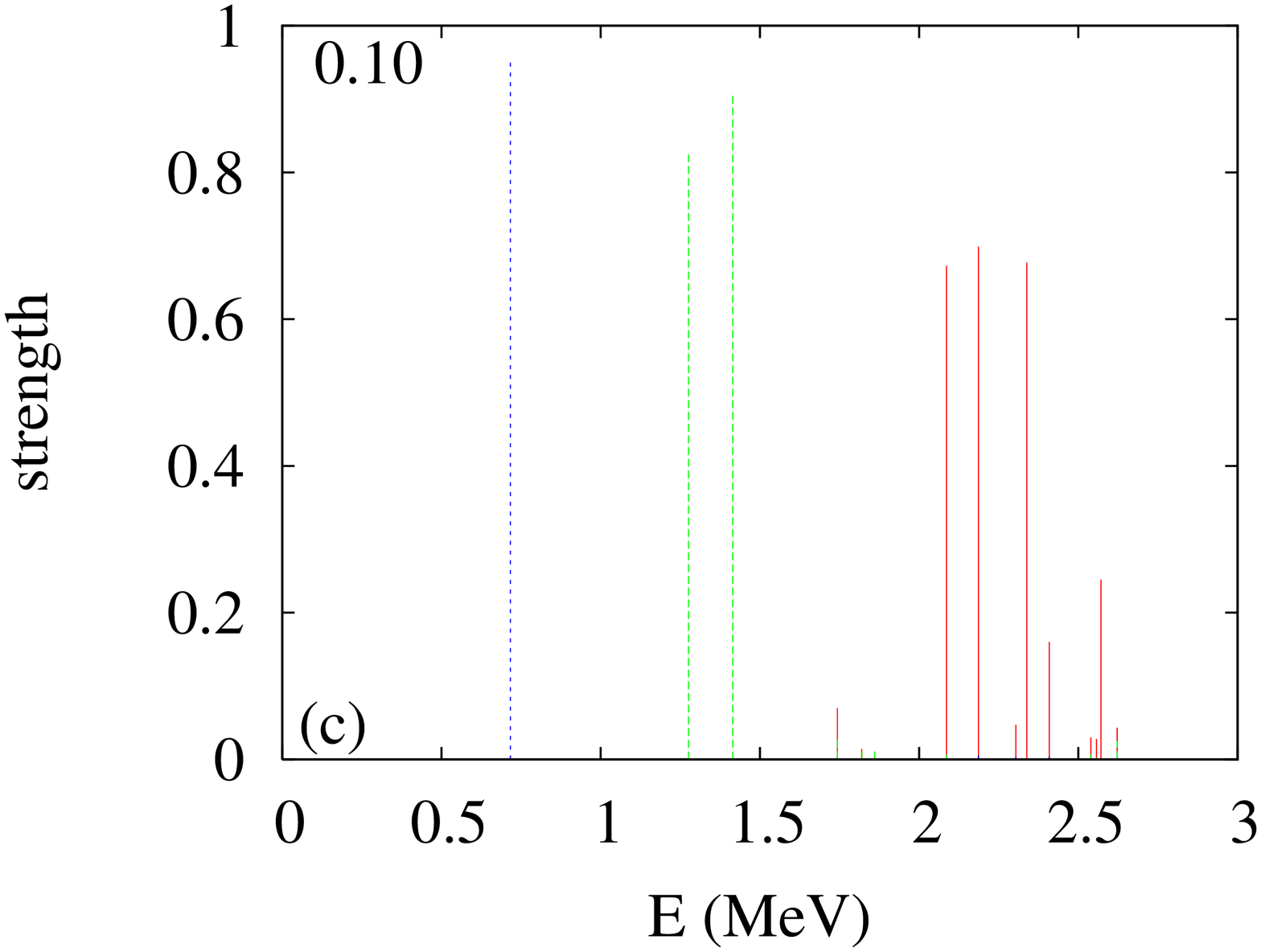}
 \includegraphics[width=5cm,angle=-90]{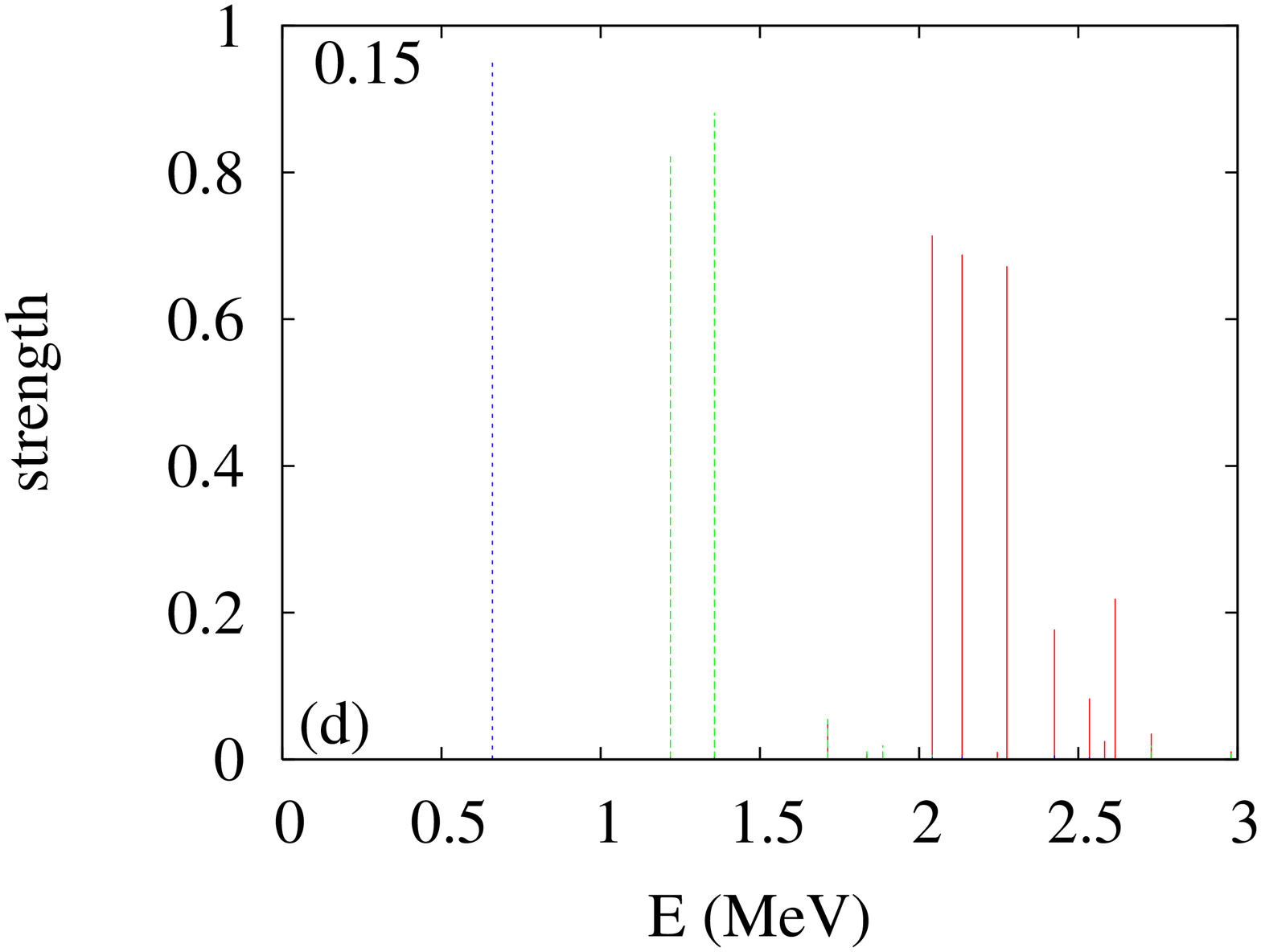}
 \includegraphics[width=5cm,angle=-90]{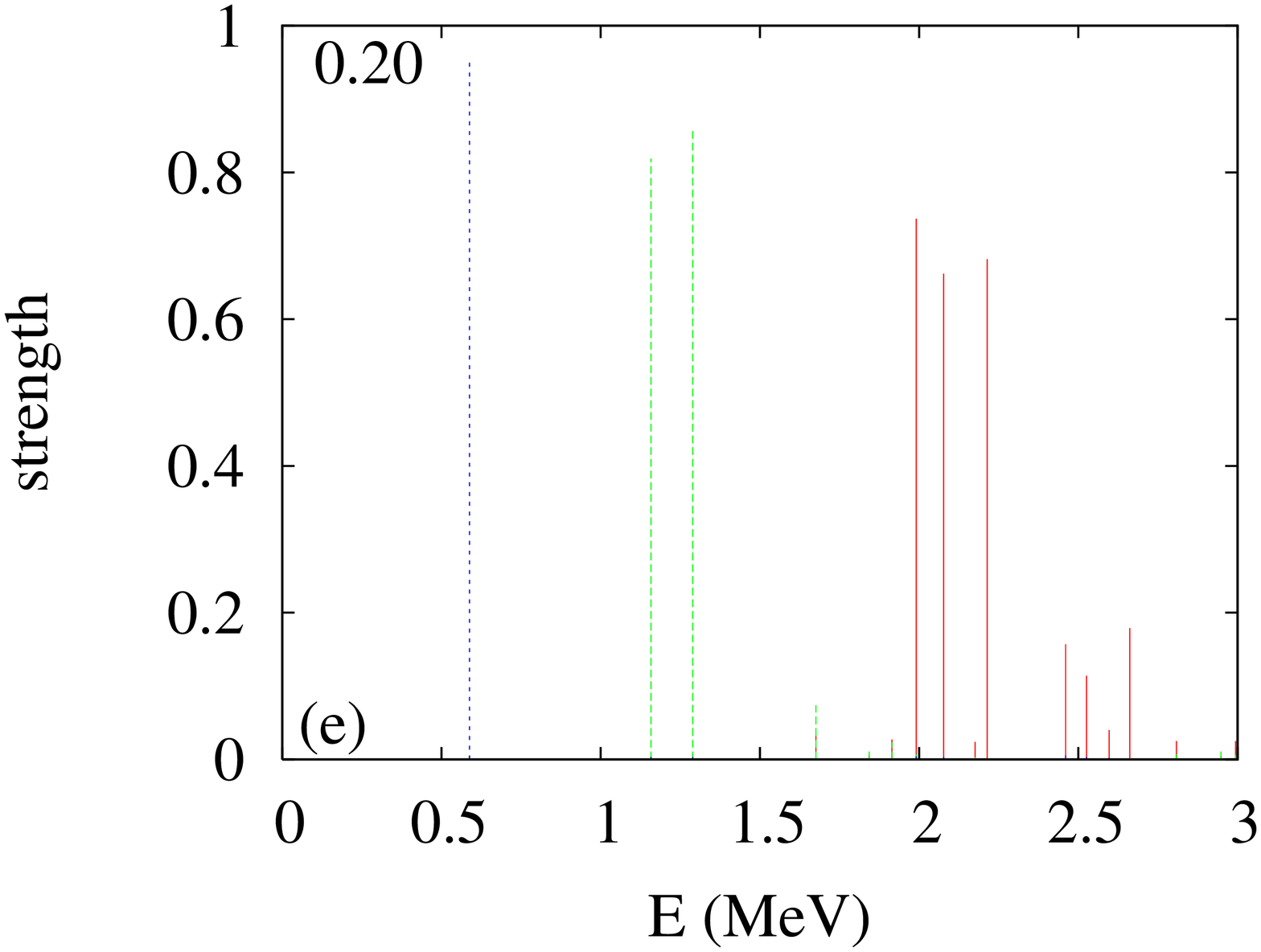}
 \includegraphics[width=5cm,angle=-90]{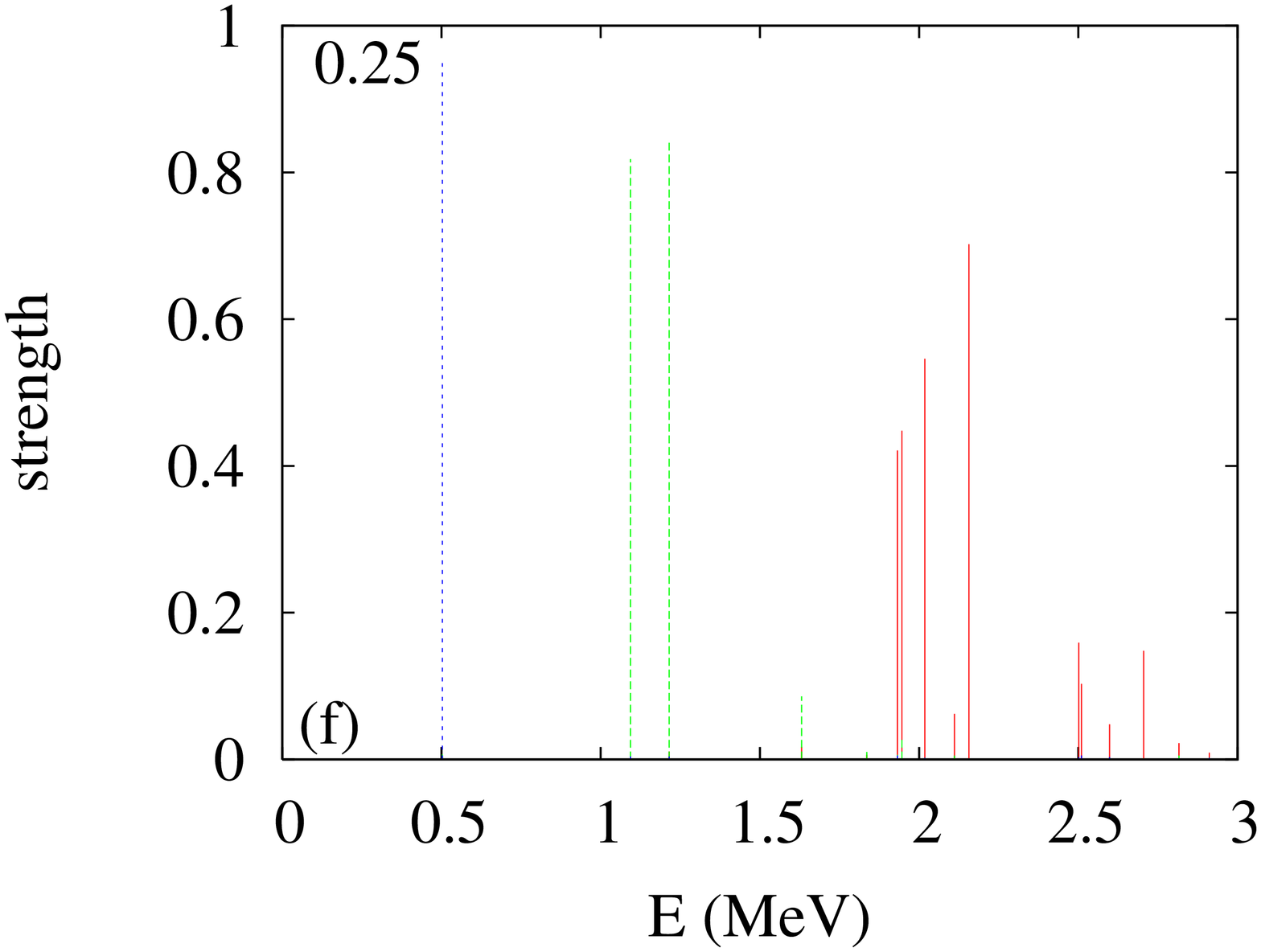}
 \includegraphics[width=5cm,angle=-90]{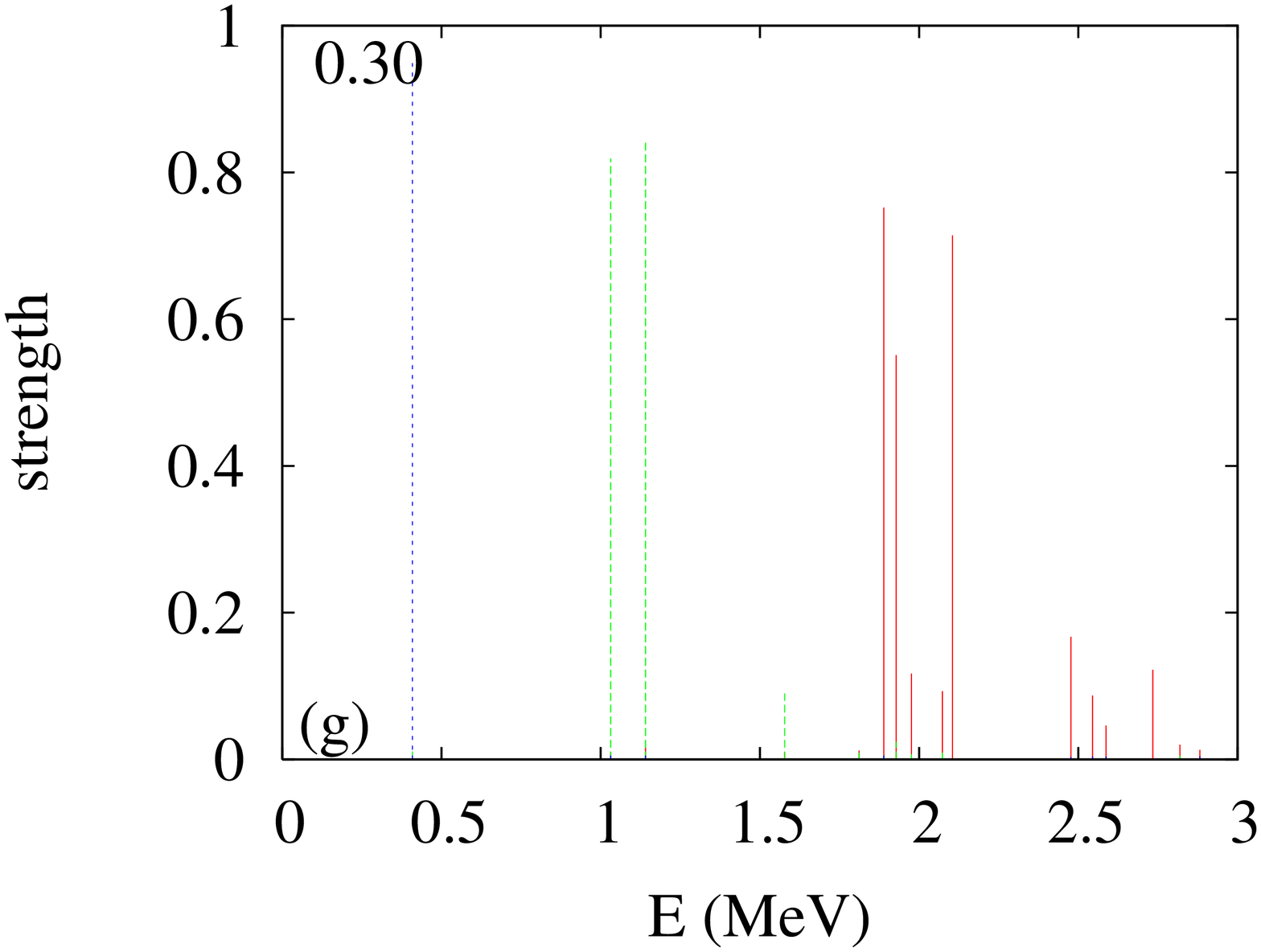}
 \includegraphics[width=5cm,angle=-90]{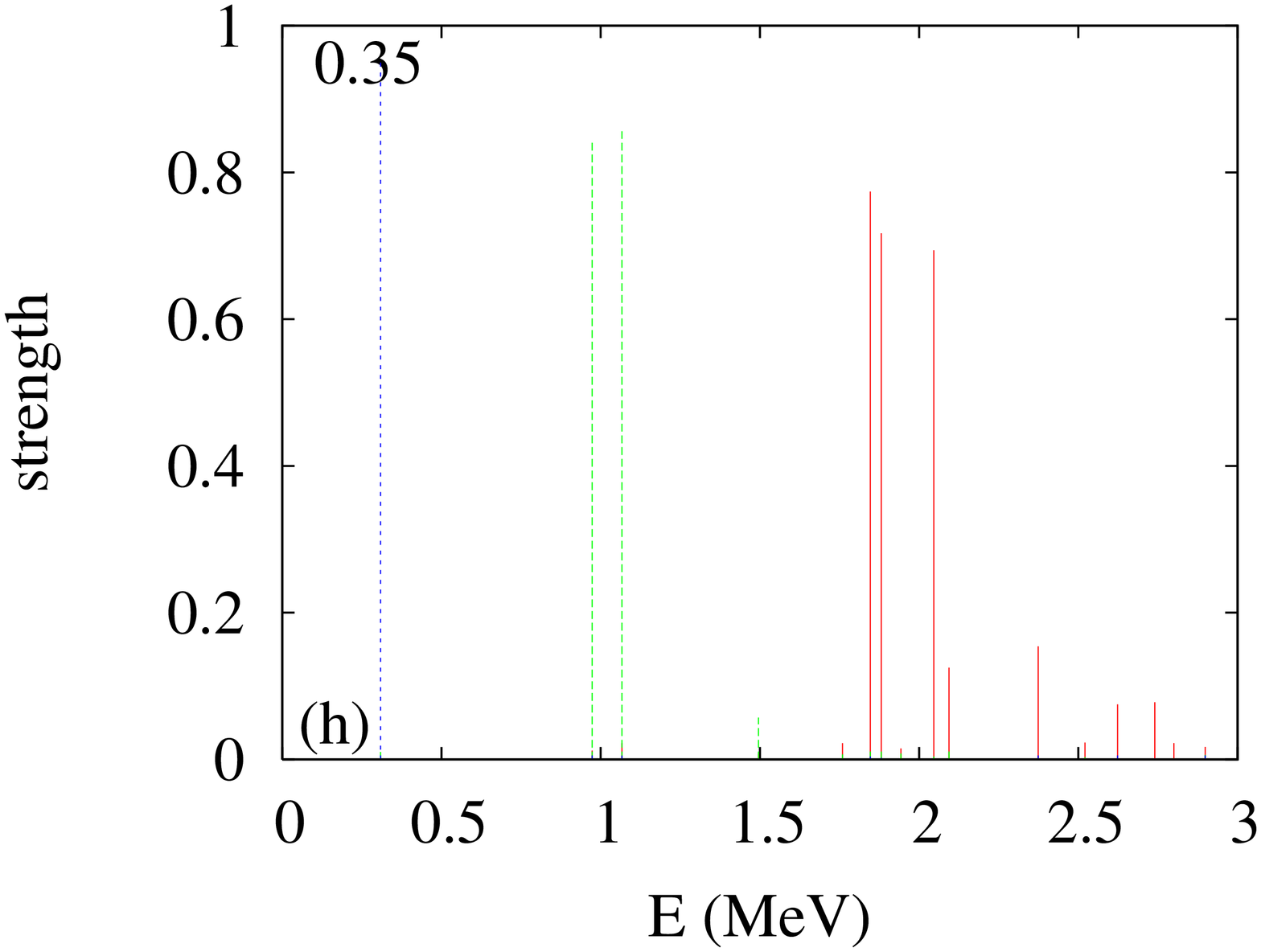}
 \caption{(Color online) Distribution of the strength (probability in the 
wave function) of the 1qp, 
1qp$\otimes\gamma$ and 1qp$\otimes\gamma\otimes\gamma$ components in the favored 
signature ($r=-i$),
$|\psi^{(1)}(\mu)|^2$ (blue dotted), 
$|\psi^{(3)}(\mu\gamma)|^2+|\psi^{(3)}(\bar\mu\bar\gamma)|^2$ (green dashed) and 
$|\psi^{(5)}(\mu\gamma\gamma)|^2+|\psi^{(5)}(\mu\bar\gamma\bar\gamma)|^2+
 |\psi^{(5)}(\bar\mu\gamma\bar\gamma)|^2$ (red solid), respectively, 
at various rotation frequencies, $\omega_\mathrm{rot}=$ 
0 - 0.35 MeV. Here 1qp ($\mu$) means the one quasiparticle states originating from 
the $\pi[422\,5/2^+]$ at $\omega_\mathrm{rot}=0$.}
 \label{fig2}
\end{figure}

Using these quantities, $H_\mathrm{couple}(\gamma)$ is diagonalized in the 
space of dimension 15 (number of quasiparticle states with $N_\mathrm{osc}$ 
= 4) $\times$ 6 (1qp, 1qp$\otimes\gamma(+)$, $\overline{\mathrm{1qp}}\otimes\gamma(-)$, 
1qp$\otimes\gamma(+)\otimes\gamma(+)$, 1qp$\otimes\gamma(-)\otimes\gamma(-)$, 
$\overline{\mathrm{1qp}}\otimes\gamma(+)\otimes\gamma(-)$, with bar 
denoting the opposite signature) for each signature sector. 
Distribution of the strength (probability in the wave function) of the 
$\pi[422\,5/2^+]\otimes\gamma$ vibration(s) that shows the collectiveness 
of each eigenstate directly is presented in Fig.~\ref{fig2} for the case 
of the favored signature $r=-i$. The result for the unfavored $r=+i$ is 
similar. 

In the favored ($r=-i$) sector, two dominantly 1$\gamma$ eigenstates are obtained 
as a result of the interaction between the $f\otimes\gamma(+)$ and the 
$u\otimes\gamma(-)$ basis states, hereafter $f$ and $u$ denote the favored and 
unfavored 1qp states originating from the $\pi[422\,5/2^+]$, respectively. 
There is no general rule 
of the correspondence between these two bands in the signature scheme and the 
$K=\Omega\pm 2$ bands in the $K$ scheme. However, since states with the lower $K$ 
have lower intrinsic energies than those with higher $K$ and the same $I$, 
the obtained lower band can be identified with the $K=\Omega-2$ band. 
Actually, in the previous analysis of the $E2$ intensity relation~\cite{Ge}, 
correspondence between the observed and the calculated states was established 
in this manner. The present result indicates that collectivity does not 
fragment much and two bands are almost parallel as in the $^{165}$Ho 
case~\cite{Ge}. But the higher band is slightly more collective and purer. 
This is consistent with the observation at $\omega_\mathrm{rot}=0$ that 
states with lower $K$ are affected more by interaction with other states~\cite{So,DP}. 
Actually only the $K=\Omega+2$ sequence was observed in many cases. 

Fragmentation of the 2$\gamma$ components is expected to depend sensitively 
on how other quasiparticle states distribute. In the present calculation, 
the 2$\gamma$ strength concentrates mainly on three eigenstates that locate at 
$\alt E_\mathrm{1qp}+2\Delta_p$. They are obtained 
as a result of the interaction among the $f\otimes\gamma(+)\otimes\gamma(+)$, 
the $f\otimes\gamma(-)\otimes\gamma(-)$ and the $u\otimes\gamma(+)\otimes\gamma(-)$ 
basis states. With the same rule as for the 1$\gamma$ case, they can be identified with 
$K=|\Omega-4|$, $\Omega$ and $\Omega+4$ from the lower. At $\omega_\mathrm{rot}=0$ 
the higher $K$ states are the more collective as expected. But small rotation 
immediately mixes their collectivity. Then three collective bands run keeping 
almost the same collectivity. 

\begin{figure}[htbp]
 \includegraphics[width=8cm,angle=-90]{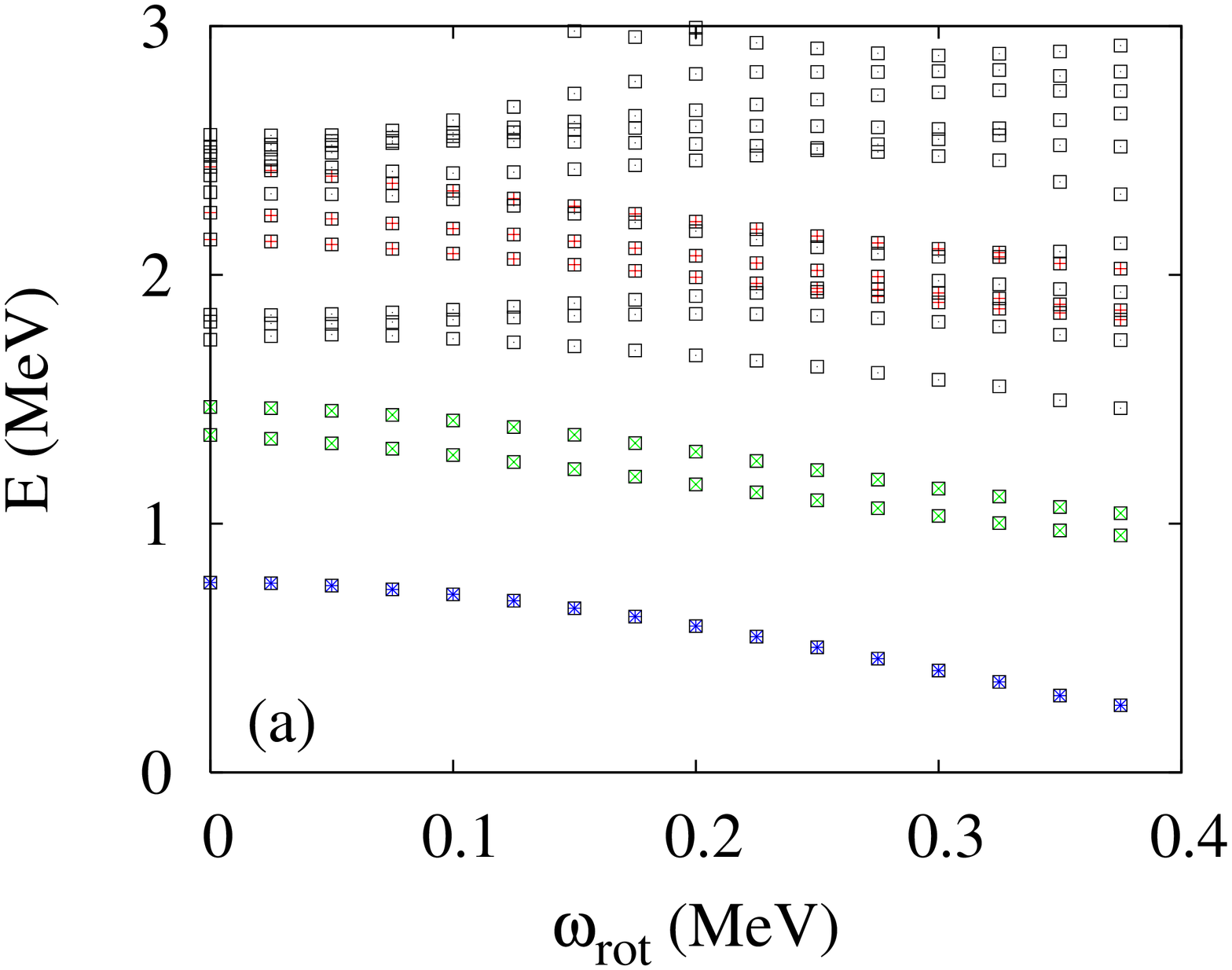}
 \includegraphics[width=8cm,angle=-90]{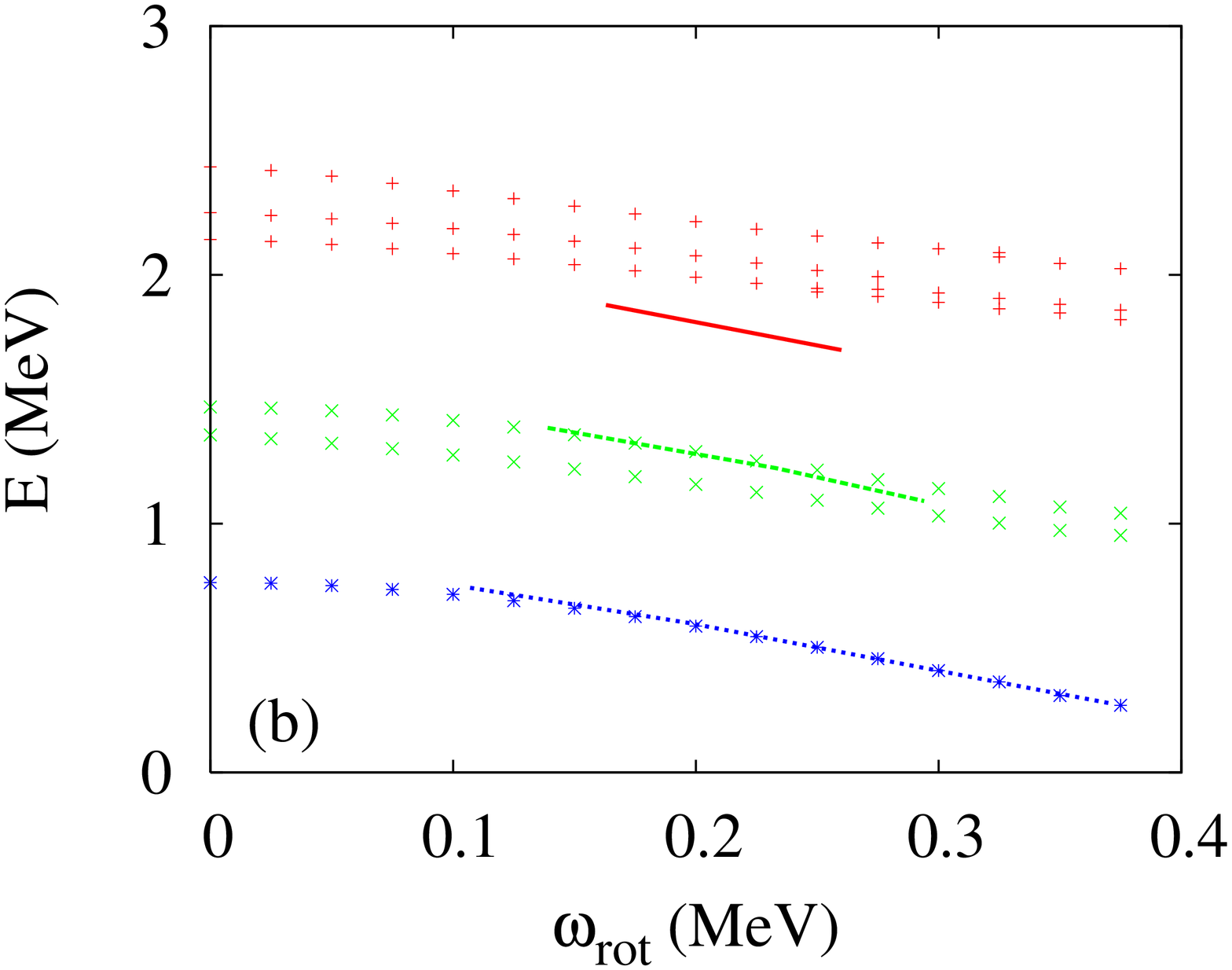}
 \caption{(Color online) (a) Calculated eigenstates of $H_\mathrm{couple}(\gamma)$ 
at each rotation frequency. Corresponding to Fig.~\ref{fig2}, those with 
large ($>40\%$) $\pi[422\,5/2^+]\otimes0,1,2\gamma$ components are marked by 
symbols. At $\omega_\mathrm{rot}=$ 0.25, 0.275 and 0.325 MeV, 
accidental fragmentation caused by 
interaction with a positive sloping 1qp state occurs. (b) Similar to (a) but 
the dominantly $\pi[422\,5/2^+]\otimes0,1,2\gamma$ states are compared with 
experimental data (curves) converted to the rotating frame by using the 
Harris parameters $\mathcal{J}_0=$ 15.45 MeV$^{-1}$ and 
$\mathcal{J}_1=$ 81.23 MeV$^{-3}$ 
that fit the yrast band of $^{104}$Mo~\cite{Mo104}.}
 \label{fig3}
\end{figure}

In order to see how these collective states interact with other non-collective 
states, all the eigenstates of $H_\mathrm{couple}(\gamma)$ that locate lower than 
3 MeV are shown in Fig.~\ref{fig3}(a). At $\omega_\mathrm{rot}=0$, there are 
2nd -- 4th $N_\mathrm{osc}=4$ 1qp states between the 1$\gamma$s and 2$\gamma$s. 
These are expected to correspond to some of the levels that were known to be 
populated by the $^{103}$Zr $\beta$ decay but spin and parity have not been 
assigned~\cite{data}. One of them interacts 
with the 2$\gamma$s depending on $\omega_\mathrm{rot}$. Aside from this, 
the 2$\gamma$s keep their collective characters as shown in Fig.~\ref{fig2}. 

Next the obtained collective states are compared with observed ones in 
Fig.~\ref{fig3}(b). This figure shows that the 1qp and the 1$\gamma$ 
($K=\Omega+2$) are reproduced almost perfectly. On the other hand, the 
calculated 2$\gamma$s obviously locate higher than the observed one. 
The most probable reason of this is ignoring higher lying states, that is, 
their effect to push down the 2$\gamma$ states. Although one may be afraid 
that inclusion of more states would lead to fragmentation of collectivity, 
we expect this not to occur since the number of nearby states is still small.

Here we discuss the adopted triaxial deformation $\gamma$ in 
Ref.~\cite{SBSP} and the present calculation. 
It has two physical aspects, its sign and its absolute value. 
The former is originally the notion in the rotating mean 
field model. Although the projection model in its original framework 
does not distinguish the sign of $\gamma$ to our knowledge, Ref.~\cite{GCS} 
proposed a method to find the main rotation axis. This made it possible to 
relate the result of the projection calculation to the sign of $\gamma$. 
In Ref.~\cite{SBSP}, the adopted $\epsilon=0.3$ and $\epsilon'=0.16$ 
correspond to $\gamma=\pm28^\circ$. 
Although its sign is not described, odd-$A$ nuclei with an $\Omega=5/2$ 
odd nucleon show $\gamma<0$ rotation due to its shape driving effect in 
addition to the approximately irrotational property of the even-even core 
in general. Then we regard the sign is consistent with ours. 
As for the latter, its absolute value, the origin of the difference 
is that of the generating mechanism of the $\gamma$ band itself as pointed 
out in Ref.~\cite{Sun}. It is constructed as a vibrational phonon excitation built 
on top of the yrast configuration in our model. In contrast, in the projection 
model, it is generated as a rotational excitation in the sense of the 
asymmetric rotor. Consequently appropriate $\gamma$ deformation is not 
necessarily the same. The latter calls for a larger $|\gamma|$; for example, 
$|\gamma|=25^\circ$ was adopted for $^{168}$Er~\cite{Sun}, which is regarded 
as a typical axially symmetric nucleus~\cite{BM}. 
Note here that large fluctuations in the $\gamma$ direction were shown for 
nuclei of that class, for example, in Ref.~\cite{Sun2}. 
These results indicate that the model dependence of the 
adopted value of $\gamma$ deformation is not unphysical. It is an interesting 
subject to look into more this longstanding problem of how these two, 
vibrational/rotational, descriptions are related, but it is beyond the scope 
of the present paper. 

Last but not least, we mention another type of collective motion for which two 
phonon excitation was observed, the wobbling motion predicted by Bohr and 
Mottelson~\cite{BM}, Janssen and Mikhailov~\cite{JM} and Marshalek~\cite{Mar}, 
and first observed by {\O}deg{\aa}rd et al.~\cite{Od}. This is a small 
amplitude fluctuation of the rotation axis of triaxially deformed nuclei 
and has the same quantum number, $r=-1$, as the odd spin members of the $\gamma$ 
vibrational bands. Two phonon wobbling bands were observed in $^{163}$Lu 
by Jensen et al.~\cite{Lu163} and $^{165}$Lu by Sch\"{o}nwa{\ss}er et 
al.~\cite{Lu165}. Their excitation energies in the rotating frame exhibit 
anharmonicity such that the two phonon states locate lower than the 
twice the energies of the one phonon states. The present author and 
Ohtsubo~\cite{MO} argued that this anharmonicity indicates the softening 
of the collective potential surface, that is, the precursor of the second 
order phase transition from the principal axis rotation to the tilted axis 
rotation. This is completely 
parallel with the situation that the softening of the $\gamma$ vibration 
is the precursor of the second order phase transition from the axial 
symmetric deformation to static triaxial deformation. The new vacuum 
after the phase transition can accommodate anharmonic vibration either 
$E_\mathrm{2-phonon}/E_\mathrm{1-phonon}>2$ or $<2$. Compare the potential 
energy surfaces: Fig.3 in the first item of Ref.~\cite{MatsuoMatsu} and Fig.3 
in Ref.~\cite{MO}; in general, stiff potentials as the former lead to high 
$E_\mathrm{2-phonon}/E_\mathrm{1-phonon}$ ratios while soft ones as the latter 
do low ratios. 

To summarize, two phonon $\gamma$ vibrational bands in rotating odd-$A$ nuclei 
were observed recently in three nuclides~\cite{Mo105,Wa,Lo}. After these 
observation, the first theoretical calculation by means of the triaxial projected 
shell model to one of them was reported~\cite{SBSP}. In the present paper, 
we have applied to the same nuclide 
a completely different, mean field based model, which was previously developed 
for the signature dependent properties of one quasiparticle (zero phonon) 
bands and later utilized for the $E2$ intensity relations in one phonon bands. 
In the present calculation, 
after confirming almost perfect reproduction of the zero and one phonon states, 
we concentrated on the effect of the odd quasiparticle 
on the distribution of the two phonon collectivity. The obtained $K=\Omega+4$ 
state is almost pure at $\omega_\mathrm{rot}=0$, but small rotation immediately 
delivers the strength to two other sequences; consequently three collective 
sequences keep about 60\% collectivity up to higher spins. 
This indicates future experimental observation of more than one sequence of the 
2$\gamma$ bands as well as the other sequence of the 1$\gamma$. 
The restriction of the model space mentioned above would be the reason 
of the result that the calculated 2$\gamma$ states locate higher than observed.  

There are some to improve the present calculation: First of all, 
the model space should be enlarged to more than two phonon states. 
Further, extension of the microscopic approach by means of the selfconsistent 
collective coordinate method~\cite{MatsuoMatsu} to rotating even and odd-$A$ 
systems is promising in future. 

\begin{acknowledgments}
The author thanks K. Matsuyanagi and Y. R. Shimizu for valuable comments. 
\end{acknowledgments}

\end{document}